\documentstyle[12pt]{article}
\newcommand{\bi}{\bibitem}
\newcommand{\be}{\begin{equation}}
\newcommand{\ee}{\end{equation}}
\newcommand{\ba}{\begin{eqnarray}}
\newcommand{\ea}{\end{eqnarray}}
\begin{document}
{\large
\begin{center}
{\bf A.A.Andrianov\footnote{
Department of Theoretical Physics,\,Institute of Physics,\,
University of Sankt-Petersburg,\,
198904,\,Sankt-Petersburg,\,Russia} and N.V.Romanenko\footnote{Department of
Neutron Research
\,, Petersburg Nuclear
Physics Institute, Gatchina, St.-Petersburg, 188350, Russia}}
\vspace{10mm}

{\Large\bf Vacuum fine tuning  and empirical estimations of masses
of the top-quark and Higgs boson}
\vspace{10mm}
\end{center}
\begin{abstract}
{\sc The fine-tuning principles are analyzed in search for
predictions of top-quark and Higgs-boson masses. The modification of
Veltman condition based on the
compensation between fermion and boson vacuum energies within the Standard
Model multiplets is proposed.
It is supplied with the  stability
under rescaling and with the requirement of minimum for the physical v. e. v.
for the Higgs field (zero anomalous dimension).
Their joint solution for top-quark and Higgs-boson couplings exists for
the cutof\/f $\Lambda \approx 2.3 \cdot 10^{13}\,GeV$ that yields
the low-energy values $m_t = 151 \pm 4\,GeV;\quad m_H = 195 \pm 7\,GeV$.}
\end{abstract}
\section{Introduction}
\qquad The standard Model(SM) describes the strong and electroweak
particle interactions with a good precision in a whole range of
energies which have been
available in experiments \cite{CERN}. Still few open problems are well known
in SM to be
resolved in order to justify all the principles which the Standard
Model is based on
and to determine eventually all its phenomenological parameters. In particular
the detection of top-quark and of  scalar Higgs particle is wanted
in the nearest future \cite{CERN,Ellis}.
Respectively the estimations for their masses have invoked a lot
of ef\/forts to understand
possible extensions of SM \cite{Ellis,Nambu} where
an underlying dynamics
leads to the formation of scalar particles \cite{Marciano,BHL}.

Meantime there exist few phenomenological principles
within the minimal SM which make it possible
to find relations between top-quark and
Higgs-boson masses and are weakly dependent of the details of a fundamental
theory underlying the SM. These principles are coming from the assumption
that the SM is actually an effective theory applicable consistently for
low energies. Accordingly its coupling constants
and dimensional parameters absorb all the influence of high-energy degrees
of freedom and of new heavy particles as well. Of course, the form of effective
action of Green-Wilson type \cite{Wil} generally depends on the
preparation procedure but the consistent ef\/fective action coinciding with
the SM action is supposedly that one which is minimally sensitive to very
high energies. Still the memory of the high-energy dynamics responsible
for the parameter formation exists just giving the relations between
dimensional parameters and certain coupling constants.
The latter statements allow to formulate following
phenomenological principles which could be fulfilled
 in the quantum SM giving various predictions for the masses of heavy
 particles.
\begin{enumerate}
\item The strong fine tuning for the Higgs field parameters
(v.e.v and its mass) that consists in the
cancellation of large radiative contributions quadratic in ultraviolet scales
bounding the particle spectra in the effective theory (Veltman condition
\noindent
\cite{Velt}).
\item The weak fine tuning that provides the cancellation of logarithmic
cutof\/f dependence in certain coupling constants and simulates thereby
the quasi-fixed ultraviolet behavior \cite{Deck-Pest,TTWu,Stech}.

Dif\/ferent implementations of both strong and weak fine-tuning we shortly
survey in Sect.2.
\noindent
\item We propose also the strong fine tuning for vacuum energies \cite{AR}
that provides the cancellation of
large divergencies quartic in ultraviolet scales which might effect
drastically in formation of cosmological constant.
Of course, a disbalance in vacuum
energies for an ef\/fective theory may happen to be compensated by those ones
from virtual high-energy components.
However we suppose here that the consistent
preparation of an ef\/fective model can provide the essential decoupling of
low-energy world from very high energies and
therefore for an appropriate choice
of ultraviolet scales the huge vacuum energies
do not naturally appear \cite{AR}.
This requirement of vacuum adaptation leads as we shall see to a
modification of Veltman condition.

The vacuum-energy fine tuning when combined with others leads to
 predictions for the top-quark and Higgs-boson masses within the range
of validity of the Standard Model.
In Sect.3 we examine the compatibility of above principles and find the
corresponding estimations for t-quark and Higgs-boson masses.
\noindent
\item The possible reduction of the SM coupling constants
\cite{Pendl-Ross,Kubo85} and
the related stability of the renormalization-group(RG) flow for the Higgs
self-coupling and/or for its coupling to t-quark (quasi-fixed infrared point
\cite{Hill}) at relatively low energies.
\noindent
\item The bounds on acceptable values of $t$-quark and $H$-boson based
on the vacuum stability in scalar sector \cite{Wein} and on the triviality
of scalar models in four dimensions \cite{Beg,triv}.
These very interesting schemes are beyond of the scope of the present
paper though
they deserve our attention when combined with the fine-tuning
rules. As well the
investigation of compatibility of fine tuning rules with scenarios of
composite Higgs-particle \cite{Rod} is postponed to a further research.
\end{enumerate}
\section{Strong and weak fine-tuning for Higgs fields}
\noindent
{\bf 1.}\quad
It is well known that the scalar sector in the Weinberg-Salam theory
contains the quadratic divergences in the tadpole diagrams and in the scalar
self-energy. In early 80th the rule of cancellation for quadratic divergences
\cite{Velt} was proposed in the  electroweak sector of the SM.
This cancellation
occurs if the fermion and boson loops are tuned due to specific
values of coupling constants. To one
loop level it was found that the condition:
\be
(2M^2_W+M^2_Z+m^2_H)=\frac{4}{3}
\sum_{flavors, \; colors} m^2_f     \label{eq:Velt}
\ee
removes quadratic divergences both from the Higgs-field v.e.v. and the
Higgs boson self-energy.
 From (\ref{eq:Velt}) it is easy to see that if the $t$-quark is the only
heavy fermion, $m_t \geq 70 GeV$ bound should hold.
\vspace{5mm}

\noindent
{\bf 2.}\quad
The idea of stability for Higgs v. e. v. against the cutof\/f variation
was extended to the weak fine-tuning in \cite{Stech}.
At the one-loop level the finiteness
of radiative corrections for Higgs v.e.v. is declared and the scale
independence of the cutof\/f $\Lambda$ is imposed.
Actually, it means the separate cancellation
of quadratic and logarithmic divergencies at one loop.
\be
\begin{array}{l}
4m_t^2=M_H^2+M_Z^2+2M_W^2;  \nonumber \\
4m_t^4=\frac{1}{2}M_H^4+M_Z^4+2M_W^4.  \label{eq:Stech}
\end{array}
\ee
The mass
predictions are: $m_t=120 GeV$; $M_H=190 GeV$. If   one takes
the value     $sin^2 \Theta_W \approx 0.25$ these conditions are
compatible  with cancellation of logarithmic divergencies in
$e^+e^- H$ vertex \cite{TTWu}(see below).  But it is worth to notice,
that the cancellation of logarithmic divergences for Higgs v.e.v
can be useful at the 1-loop level only if
two-loop quadratic divergencies are smaller than the 1-loop
logarithmic ones. But for the values of $\Lambda \ge 10 TeV$
logarithmically divergent part in v.e.v is less than $10^{-3}$
in comparison with 1-loop quadratic divergencies
(the factor  $\frac{v^2}{\Lambda^2}\ln\frac{\Lambda^2}{v^2}$ )
while two-loop quadratic corrections have the factor $10^{-2}$
 ($1/16 \pi^2$). It means that even at smaller energies two-loop
 contributions cannot be neglected (we shall illustrate
 this fact more carefully further on). One should also
bear in mind that for energies less or equal than 1 TeV the absolute
value of the logarithmic contributions to v.e.v is rather small in
comparison to the bare v. e. v. and that in the SM
there are quite a few of logarithmic
divergencies which  survive after this cancellation (see below).
\vspace{5mm}

\noindent
{\bf 3.}\quad The further development of above tuning
was outlined by R.Dec\-ker and J.Pesteau
\cite{Deck-Pest}: their idea (in 1980) was of the lepton mass finiteness
(based on the assumption that leptons are not composite
particles even beyond the SM). Their one-loop calculation deals with
logarithmic fine-tuning including the logarithms from Higgs tadpoles and
leads to the following relation:
\be
2M_W^4+M_Z^4+\frac{1}{2}m_H^4+2M_W^2\cdot  \sin^2 \theta_W \cdot
m_H^2=\frac{4}{3}
\sum_{flavors,\;colors} m_f^4 \;  .  \label{eq:2}
\ee
 This relation yields the lower mass bound for the $t$-quark $m_t\geq 78GeV$,
 while combined with the $\rho$-parameter restriction \hfill
 $m_t \leq \\ 200 GeV$
 yields also the upper bound $M_H \leq 350 GeV$  (in 1980 it was
 $m_t \leq 295 GeV$ , $M_H<450GeV$ respectively).

The contribution of the Higgs-boson exchange
to the fermion self-energy
 is neglected in (\ref{eq:2}) since all the leptons are
  light compared with $M_{W,Z}$ and thereby the  self-energy of
$t$-quark (and of other heavy quarks) remains
  large. This is the reason why the lepton self-energy only
has been declared to be stabilized.
On the other hand in the approach\cite{Deck-Pest}
the quadratic divergences of tadpoles
were not taken into account (in their usage of dimensional regularization
  the d=4 poles are equivalent to logarithmic divergences only).
However the equation (\ref{eq:2}) is incompatible with (\ref{eq:Velt}),
when one cannot avoid the problem of
  quadratic divergences ignored in (\ref{eq:2}).

Another application of this scheme has been
made recently providing cancellation in the neutrino self-energy and leading
to the equation:
\be
2M_W^4+M_Z^4+\frac{1}{2}m_H^4+\frac{m_e^2-m_{\nu_e}^2}{2}m_H^2=\frac{4}{3}
\sum_{flavors,\;colors} m_f^4 \;  .  \label{eq:2a}
\ee
For the neutrino mass the quadratic and logarithmic
divergences can be cancelled
simultaneously leading to the prediction: $m_t=121\,GeV$, $M_H=194\,GeV$.

Still one can dispute  with the very idea to compensate
large logarthmic corrections coming from tadpoles non-running under
renormalization and from self-energy diagrams contributing
into the anomalous dimension of the mass. The former ones provide the
finiteness of the basic EW scale when being cancelled in the Higgs-field
v. e. v. (see (\ref{eq:Stech})). If it takes place then one can
easily check that neither (\ref{eq:2})
nor (\ref{eq:2a}) are not fulfilled simultaneously.
\vspace{5mm}

\noindent
{\bf 4.}\quad
The weak fine tuning has been applied to the $e^+e^-H$ vertex
in \cite{TTWu} as the cancellation of
 logarithmic divergences therein. This vertex is not af\/flicted with
quadratic divergences and therefore it is safe in the weak fine
tuning.The equation dif\/ferent of (\ref{eq:2}),(\ref{eq:2a}) has been
obtained,
 \be
m_t^2= \frac{5}{2}M_Z^2-M_W^2 \label{eq:4}
\ee
which is well compatible with (\ref{eq:Velt}), and yields
the following mass predictions for $m_t=120GeV$, $M_H=190GeV$.
\vspace{5mm}

\noindent
{\bf 5.}\quad
Few papers were in search of certain unknown
  symmetries within or beyond the SM
which would provide the systematic cancellation of the quadratic
  divergencies. If this symmetry were local then quadratic divergencies
  would cancel at each order of the perturbation theory.
  The special continuation of the dimensional regularization method
 for two-loop quadratic divergencies aside $dim= 4$ was adopted in \cite{Ruiz}.
  However,  methods used in \cite{Ruiz}-\cite{AntiRuiz}
  treat poles in $d\not=4$ at dif\/ferent positions for a particular number of
loops and bring different results dependimg on the recipe of contituation
in dimensions. The dimensional reduction yields the one-loop
Veltman condition, whereas the dimensional regularization
replaces it by another equation, whose physical meaning is obscure
(see the
discussion in \cite{Einhorn}).
Furthermore the separate cancellation
  of one and two-loop condition happens to be compatible
only if the QCD coupling is actually turned off.
  The possibility of three-loop cancellation (which is necessary
  under such a treatment) remains an open question \cite{Jack}.

\vspace{5mm}
\noindent
{\bf 6.}  \quad
  Another interesting discovery had been brought by computing
of the RG-flow for coupling $\lambda$ (the 1-loop equation for $\lambda$
  is nonlinear and cannot be solved exactly):
\be
16\pi^2 \frac{\partial \lambda}{\partial \tau}=
12 \left( \lambda^2 +(g_t^2-A) \lambda -g_t^4 +B \right)    \label{eq:9}
\ee
$$ A= \frac{1}{4}\left(3g^2+g'^2 \right) \; ; \;
B=\frac{1}{16}g'^4+\frac{1}{8}g^2g'^2 +\frac{3}{16}g^2 $$
It was found \cite{BHL,Hill} that $\lambda(\tau)$ tends to
the Hill's quasi-fixed point for $\lambda$ ($\partial \lambda/
\partial \tau=0$) in the wide intermediate energy region for
\underline{any} boundary conditions at high energies.
For their composite H-boson scenario this yields
$m_t\approx 240GeV,\quad M_H\approx 250 GeV$,
but the above property of eq.(~\ref{eq:9}) should be taken into
account for any low-energy predictions in the SM.

Thus we shall follow the preparation way for a low-energy ef\/fective
action  based on a momentum cutof\/f.
\section{ Vacuum-energy fine tuning and predictions for $t$-quark and
$H$-boson masses}
Let us consider the SM as a low-energy limit of a more
fundamental theory and suppose that the only heavy fermion, t-quark
is involved in its dynamics within the selected energy range. Respectively we
neglect the masses of all lighter fermions. When there is no expected
supersymmetry below Grand-Unification scales we apply different
scales for the design of SM-effective action for bosons $\Lambda_{B}<<
\Lambda_{comp}$ and for fermions $\Lambda_{F}<< \Lambda_{comp}$. Among
bosons the universal scale is introduced in order not to induce the
explicit breaking of a Grand
Unification symmetry below a scale of compositeness $\Lambda_{c}$. As well
the unique scale for fermions ensures the horizontal symmetry in the
ultraviolet region.

We require for the SM the suppression of very large contributions (leading
divergencies) into dimensional physical parameters
that is equivalent to the absence  of their strong scale dependence.
The latter means in addition to the strong fine-tuning that
the cancellation of contributions into vacuum energy should take place,
i.e., first of all the contributions which are quartic
in cutof\/fs.
\begin{center}
\ba
T_{\mu \nu}\sim g_{\mu \nu}E_{vac}\approx 0;\quad
4N_F\,\Lambda_F^4 = (3N_B + N_S)\,\Lambda_B^4;\nonumber\\
\alpha^2=\frac{\Lambda_b^4}{\Lambda_F^4}=\frac{4N_F}{3N_B+N_S} = 2.4
\ea
\end{center}
where $N_F = 24$ is a number of flavor and color fermion degrees of
freedom for three generations, $N_B = 12,\, N_S= 4$ are numbers of
flavor and color degrees of freedom for vector and scalar bosons
respectively.

The strong fine-tuning condition in this case reads at the one-loop
level
\be
4m^2_t=\alpha(2M^2_W+M^2_Z+m^2_H)
\ee
 Taking into account the ef\/fects of all loops one comes to the integral
Veltman condition:
\ba
\int_{v_0}^{\Lambda_{F}}\frac{4g_t^2 d^4k}{k^2+m_t^2}=
\int_{v_0}^{\Lambda_{B}}d^4k\biggl\lbrace\frac{g^2}{k^2+M_W^2}+ \nonumber \\
\frac{(g^2/2 +g'^2/2)}{k^2+M_Z^2} +
\frac{\lambda}{k^2}   +
\frac{\lambda}{k^2+ M_H^2}\biggr\rbrace
\ea
where the conventional denotations for the electroweak $g,\,g'$, Higgs-quartic,
$\lambda$ and the Yukawa $t$-quark, $g_t$ coupling constants are used.
When integrating by  parts one can conclude that the leading contribution is
the modified Veltman condition at the scale $\Lambda$. The latter is
supplemented in the next-to-one-loop approximation
with its renorm-derivative (having small combinatorial factor) and so on.
At one-loop level of accuracy, all the renorm-derivatives
except for the first one are zero.
Demanding the weak dependence of $\lambda$ one has
to impose both  the modified Veltman condition and its renorm-derivative.

\ba
D \equiv 16 \pi^2 \frac{\partial}{\partial \tau};\quad
\tau = \ln\frac{\Lambda}{v_0}\nonumber\\
\left\{
\begin{array}{r}
f \equiv 4 g_t^2 - 2 \alpha (\lambda +A)=0  \\
Df=0
\end{array} \right.
\ea
The explicit form of the second stability condition is:
\begin{eqnarray}
Df  = & 8g_t^2 \left[ \frac{9}{2}g_t^2-8g_3^2-\frac{9}{4}g^2-
\frac{17}{12}g'^2 \right] & \nonumber \\
   &     - 24 \alpha \left[ \lambda^2 + (g_t^2-A)\lambda +B - g_t^4 \right] &
   \nonumber \\
  & - 2 \alpha (-19g^4 + \frac{41}{3}g'^4)/4 &
   \end{eqnarray}
where in Eqs.(10), (11) the denotations are borrowed from \cite{Hill}:
\be
A \equiv \frac{3}{4}g^2+\frac{1}{4}g'^2;\quad
B=\frac{1}{16}g'^4+\frac{1}{8}g^2g'^2 +\frac{3}{16}g^2
\ee
In order to calculate the solution of Eqs.(10) let us introduce the following
variables:
\be
x_A \equiv g_t^2/A;\quad
y_A \equiv \lambda /A;\quad
z_A \equiv g_3^2/A
\ee
Evidently,
\be
y_A = \frac{2}{\alpha} x_A - 1.
\ee
Then the equation for the $t$-quark Yukawa coupling constant reads,
\be
k_1 x_A^2 +k_2 x_A + C =0 ,
\ee
where the coef\/f\/icients are:
\begin{displaymath}
k_1=36-24 \alpha(4/ \alpha^2 +2/ \alpha -1) \approx -36.787;
\end{displaymath}
(it does not depend on energy scale)
\ba
k_2= -64 z_A -24 - \frac{16}{3} \frac{g'^2}{A} +24 \alpha(\frac{6}{\alpha}
     +1) \nonumber\\
C=-24\alpha(1+\frac{B}{A^2}) - \frac{2\alpha}{A^2}(-19g^2+\frac{41}{3}g'^2)/4
\ea
Numerically the existence of a solution is very sensitive to both the
value of $\alpha\simeq 1.55$ and the value of the strong coupling constant
$\alpha_3 = g^2_3/4\pi$. In what follows the updated averaged value
of $\alpha_3$ is taken from \cite{Alt} as
$ \alpha_3(M_Z) = 0.118 \pm 0.007$.

In Table 1 one can find the estimations for masses of $t$-quark and $H$-boson
for dif\/ferent cutof\/fs in preparation of low-energy ef\/fective action
for SM.
\vspace{5mm}

\begin{center}
\begin{tabular}{||l|l |lr|r||}                                          \hline
$\Lambda$, GeV  &   $10^{15}$ & \multicolumn{2}{c||}{$10^{14}$} &
$2.3\cdot 10^{13}$\\ \hline
$\ln \Lambda $  &  34.54    &  \multicolumn{2}{c|}{32.24} &  30.63         \\
$\alpha_3(\Lambda)$&  0.0233     &  \multicolumn{2}{c|}{0.0248} & 0.0261
\\
$\alpha_2(\Lambda)$&  0.0222     &  \multicolumn{2}{c|}{0.0228}&0.0231
\\
$\alpha_1(\Lambda)$&  0.0138     &  \multicolumn{2}{c|}{0.0135}&0.01335
\\
$m_t(\Lambda)$, GeV &  105      &  99      &     79             & 89 \\
$M_H(\Lambda)$, GeV &  114      &  98      &     22             & 68   \\
$m_t$(100 GeV)      & 170      & 164      &  146                 & 155 \\
$M_H$(100 GeV)      & 202      & 196      &  177                 & 187 \\
\hline
\end{tabular}

\vspace{10mm}
{\bf Table 1.}\quad Predictions from modified Veltman condition and its
renorm-derivative.
\end{center}
\vspace{5mm}
The low energy values of $m_H$ are evaluated with help of the IR quasi-fixed
point for the Higgs self-coupling \cite{Hill} which has been established
at one-loop level.

The above stability conditions ensure the strong f\/ine-tuning both to
two-loop level and numerically. So one may consider the Higgs-f\/ield v.e.v.
as a fundamental scale of SM.
Still the physical parameter $v$ related to
the above v.e.v. appears in the renormalized lagrangian in place of
$\langle H \rangle \simeq v_0$ and dif\/fers from it due to the wave function
renormalization. The latter dif\/ference arises from non-vacuum diagrams
and creates the anomalous dimension for the renormalized value of $v$.

The renormalization-group equation then deals with the RG flow generated by
\be
\begin{array}{l}
\gamma_v= \left( \frac{3}{4} \left(3g^2 + g'^2 \right) - 3g_t^2 \right) v(M).
\end{array}
\ee
The natural supplement for the set of stability conditions of Veltman type
might be the requirement to have zero anomalous dimension for $v$ that in
turn corresponds to the true minimum for the spontaneous symmetry-breaking
ef\/fects \cite{Stech}. Luckily it happens to be compatible with modified
Veltman conditions. Below on we display the joint solution of equations
$\gamma_v = 0;\, f = 0$ for a wide range of cutof\/fs.

\vspace{5mm}
\begin{center}
\begin{tabular}{||l|l|l|r||} \hline
$\Lambda$, GeV  &  $10^3$ & $10^4$ & $10^5$          \\ \hline
$A=g_t^2$ &   0.34485 & 0.33345 & 0.32196  \\
$m_t(\Lambda)$, GeV&   102.1 & 100.4 &  98.7   \\
$M_H(\Lambda)$, Gev &  77.8  & 76.9   &  75.2   \\
 $\frac{v^2}{\Lambda^2}\ln \frac{\Lambda^2}{v^2}$ & 0.106
& $2.45*10^{-3}$& $3.85*10^{-5}$\\ \hline
 \end{tabular}
\vspace{3mm}

\begin{tabular}{||l|l|l|r||}  \hline
$\Lambda$, GeV  &     $10^6 $ & $10^7$  & $10^8$        \\ \hline
 $\alpha_2(\Lambda)$&0.02935  & 0.02834 & 0.02739        \\
$\alpha_1(\Lambda)$&0.01166   & 0.01187 & 0.01208        \\
$A=g_t^2$          & 0.3133   & 0.3044  & 0.2961         \\
$m_t(\Lambda)$, GeV& 98.0     & 96.5     & 95.2          \\
$M_H(\Lambda)$, GeV& 74.7     & 73.7     & 72.6          \\ \hline
\end{tabular}
\vspace{3mm}

\begin{tabular}{||l|l|l|r||}  \hline
$\Lambda$, GeV  & $10^{9}$ & $10^{10}$ &  $10^{11}$   \\ \hline
 $\alpha_2(\Lambda)$& 0.02650 &0.02567 & 0.02489       \\
$\alpha_1(\Lambda)$& 0.01229  & 0.01252 & 0.01275       \\
$A=g_t^2$          & 0.2884 & 0.2813   & 0.2747         \\
$m_t(\Lambda)$, GeV& 94.0   & 92.8     & 91.7           \\

$M_H(\Lambda)$, GeV& 71.7   & 70.8     & 70.0           \\ \hline
\end{tabular}
\vspace{3mm}

\begin{tabular}{||l|l|l|l|r||}  \hline
$\Lambda$, GeV  & $10^{12}$& $10^{13}$&$10^{14}$&$10^{15}$   \\ \hline
 $\alpha_2(\Lambda)$&0.02456 &0.02346& 0.02281& 0.02219       \\
$\alpha_1(\Lambda)$& 0.012997& 0.01329& 0.013512& 0.01378   \\
$m_t(\Lambda)$, GeV& 90.2  & 89.2      &  88.3  & 87.9     \\
$M_H(\Lambda)$, GeV& 69.2  & 68.0      &67.3     &  67.1    \\ \hline
\end{tabular}
\vspace{5mm}

{\bf Table 2.} Predictions from the modified Veltman condition and
the absence of anomalous dimension.
\end{center}
\vspace{5mm}

One can compare Tabs. 1 and 2 and find the overlapping for
$\Lambda \sim 2.5\cdot10^{13} GeV$  that gives in turn
$$m_t(100GeV) \approx 155GeV;\, m_H(100GeV) \approx 187GeV$$.
On the other hand the usage of one-loop approach when deriving low-energy
values brings a theoretical error that can be estimated by averaging
of discrepancy between modified Veltman condition and Hill's condition
as follows:
\ba
\left\{
\begin{array}{l}
m_t = 151 \pm 4\, GeV\\
m_H = 195 \pm 7\, GeV
\end{array} \right.
\ea
These predictions are within the accepted range for above masses
found by overlapping of different experimental and theoretical bounds
\cite{Ellis}.
\section{Conclusions and extensions}
\qquad We have shown that the modification of Veltman condition caused
by the strong fine-tuning of vacuum energies makes it possible to define the
ef\/fective SM with a finite cutof\/f  with fundamental EW scale
and related particle masses essentially less than the cutof\/f.
It has not been available in the
original formulation (for $\alpha = 1$ there is no solution \cite{Jack}).
Furthermore
it is consistent also at the two-loop level and compatible with the absence
of anomalous dimension for the electroweak scale.
The corresponding  mass predictions are:
$m_t\approx 151 \pm 4 \, GeV$;  $M_H\approx 195 \pm 7\,GeV$. The
value of the cutof\/f contains an uncertainty
connected with the experimental error in $\alpha_S$,
the latter one causes the following error bar for cutof\/f,
$5\cdot 10^{12}\,GeV<\Lambda_F< 5\cdot 10^{14}\,GeV$.
However, the masses can be defined more accurately through the absence
of the anomalous dimension and the RG flow.

It may be  interesting
to combine the above conditions with other principles following from the
infrared analysis. In particular we pay attention to the embedding the vacuum
fine tuning into models with composite scalars following the discussion in
\cite{Rod}. It definitely will bring the intermediate scale where the
above fine-tuning relations are necessary conditions to start the
renormalization-group flow down to lower energies for
all dimensional parameters with the RG equations of the conventional SM.
\section{Acknowledgements}
One of us (A.A.) is very grateful to the III.Physikalische Institut,
RWTH - Aachen and especially to Prof. Dr. R. Rodenberg for hospitality and
fruitful discussions and to the DFG for financial support.

}
\end{document}